\documentclass[12pt, a4paper]{article}
\usepackage[margin=2.5cm]{geometry}
\usepackage[english]{babel}
\usepackage[utf8]{inputenc}
\usepackage[T1]{fontenc}
\usepackage{titling} 

\usepackage{graphicx}
\graphicspath{{./Images/}}
\usepackage[bf]{caption} 

\usepackage{setspace}

\usepackage[square, numbers, sort&compress]{natbib} 

\usepackage{amsthm,amsmath,amssymb,bbm}
\usepackage{tensor,mathtools,slashed,braket}
\usepackage{float,wrapfig} 
\usepackage[compat=1.1.0]{tikz-feynman} 
\tikzfeynmanset{warn luatex = false} 

\usepackage{hyperref}

\newcommand{\mlc}[1]{} 
\newcommand{\email}[1]{\thanks{Electronic Address: \href{mailto:#1}{#1}}}

\newcommand{\p}[1]{(\ref{#1})}
\newcommand{\be}{\begin{equation}}
\newcommand{\ee}{\end{equation}}
\newcommand{\bea}{\begin{eqnarray}}
\newcommand{\eea}{\end{eqnarray}}

\newcommand{\Vgr}{V_{3\text{gr.}}}
\newcommand{\Vgri}[1]{V_{3\text{gr.},#1}}
\newcommand{\A}[2]{A_{#1#2}}
\newcommand{\Pa}[2]{P_{#1#2}}

\begin{document}
	\title{BRST methods for constructing quartic actions for spinning black holes 
    }
	
	\preauthor{\begin{center}
			\large \lineskip 0.5em%
			\begin{tabular}[t]{c}}
			
			\author{Julian Lang\email{julian.lang.research@gmail.com} ~and Mirian Tsulaia\email{mirian.tsulaia@gmail.com}}
			
			\postauthor{\vspace{0.5em}\\
				\textit{Okinawa Institute of Science and Technology},\\
				\textit{1919-1 Tancha, Onna-son, Okinawa 904-0495, Japan}%
			\end{tabular}\par\end{center}}
	
	\date{\today}
	
	\begin{titlingpage}
		\maketitle
		
        \begin{abstract}
			We develop a systematic approach to the computation
            of gauge invariant quartic interactions
            between  reducible massive and massless higher spin fields. Extending the BRST formulation of existing cubic results, we obtain a single constraint for each off-shell quartic vertex that ensures both the gauge invariance of the Lagrangian and formal closure of the gauge algebra at quartic order. A solution to these equations is presented. The general equation is then reduced to an on-shell version to reduce complexity. We find example solutions for the off-shell and on-shell quartic vertices in low spin examples relevant to the problem of black hole scattering.
		\end{abstract}
	\end{titlingpage}
	
	\tableofcontents
	\newpage

	\onehalfspacing
	
	\section{Introduction}


    Since the first direct observation of Gravitational Waves was made a decade ago, there has been growing interest in the study of their properties. The system of particular interest to us is the system of Rotating (Kerr) black hole binaries interacting via the exchange of Gravitational Waves.
    
    A possible way to describe such a system is an effective field theoretic approach in the Post Minkowskian approximation~\cite{Cangemi:2022bew}.
    It was found that the corresponding effective field
    theory is one that contains massive higher spin fields, 
    interacting via a massless graviton.
    In particular, this connection can be demonstrated
    by taking the classical limit
    of the three point function \cite{ArkaniHamed2017}
    between two fields
    with spin $s$ and mass $m$ interacting with a graviton
    \be \label{eq:intro:min-c-2}
        M_3^{s,s,2^+} =  M_3(0,0,2^+)
        \frac{\langle 2 1 \rangle^{2s}}{m^{2s}}\ , 
    \ee
    where $M_3(0,0,2^+)$ is a minimal coupling
    of a scalar field to a helicity $2$ massless boson.
    The classical limit
    \cite{Guevara:2018wpp,
    Chung:2018kqs, Chiodaroli:2021eug}
    gives exactly the  interaction vertex
    of the Kerr black hole with the massless graviton
    in the Post Minkowskian approximation \cite{Vines:2017hyw}.

    A natural question that arises is how to
    construct a field theory Lagrangian which gives the coupling
    \p{eq:intro:min-c-2}, when the fields are taken to be in- and out-
    on-shell states. A possible solution was suggested
    in \cite{Skvortsov2024} in terms of the gauge invariant
    massive higher spin field theory
    \footnote{ Worldline actions for Kerr black holes were recently suggested in 
   \cite{ Kim:2025xka, Ben-Shahar:2025tiz}.},
    which will be referred to as the BRST approach in the following discussion.\footnote{See
    \cite{Lang:2025ifq} for a short review of the BRST approach to massive higher spin gauge theories
    and their application to the description of spinning black hole binaries.
    Gauge invariant formulations of free massive 
    higher spin fields can be found in \cite{Pashnev:1989gm, 
Pashnev:1997rm,
Klishevich:1997pd, Bekaert:2003uc,
Buchbinder:2005ua,
Francia:2010qp, Chekmenev:2019ayr,
Fegebank:2024yft,
Lindwasser:2023zwo}. 
}
This approach closely parallels
the one of bosonic Open String Field Theory\footnote{Three point amplitudes that correspond to root-Kerr solutions in the context of string theory were discussed in \cite{Alessio:2025nzd}.}
(OSFT)
and shares several properties with it.
First, it is a gauge invariant formulation of massive and massless higher spin fields and therefore contains physical and auxiliary fields. Second, 
one can choose to consider reducible representations of the Poincar\'e group, instead of irreducible ones. Despite reducible spin $s$ fields containing a tower of irreducible components of spins $s,s-2,..,1/0$ (like daughter Regge trajectories in String Theory) the cubic and higher-point interactions actually become much simpler. We will therefore be dealing with reducible fields.
Let us note that unlike OSFT one can construct a nilpotent BRST charge and thus obtain a consistent gauge invariant free Lagrangian for massive and massless higher spin fields in any number of space-time dimensions.
 
As far as cubic interactions are concerned, again similar to OSFT \cite{Neveu:1986mv, Gross:1986ia},
the gauge invariance requires the cubic vertices to be BRST invariant
(see  \cite{Fotopoulos:2008ka} for a review).
A complete classification of BRST invariant cubic vertices for all possible combinations of 
masses and spins is given in \cite{Metsaev:2012uy}\footnote{The classification in the light cone approach was given in
\cite{Metsaev:2005ar,Metsaev:2007rn,Metsaev:2022yvb}.
Cubic vertices for massless fields, when the corresponding BRST charge describes irreducible representations of the Poincar\'e group were constructed in \cite{Buchbinder:2021xbk}.} 
The building blocks for the cubic vertices
are BRST invariant expressions, and any function
of these expressions is a priori (i.e., if one is concerned only with cubic interactions) a valid cubic vertex.
One can show \cite{Skvortsov2024}, that
taking a particular function
of these BRST invariant expressions
gives us the three point function \p{eq:intro:min-c-2} (known also as the minimal coupling)
after going on-shell and using the four dimensional
 spinor-helicity formalism \cite{Dittmaier:1998nn}--\cite{Conde:2016izb} \cite{ArkaniHamed2017},
 for massive and massless fields.
Similarly by appropriately choosing
a different function of the BRST invariant expressions
one can arrive at the three point function
for minimally coupled interaction of
a massive complex field with an arbitrary spin and a massless vector field 
\cite{ArkaniHamed2017}
\be
M_3^{s,s,1^+} =  M_3(0,0,1^+) 
 \frac{\langle 2 1 \rangle^{2s}}{m^{2s}}\ .
\ee

The goal of the present paper is to extend the analysis
of \cite{Skvortsov2024} to the level of quartic interactions. To this end, we again use
the BRST approach with cubic vertices 
that contain two copies of a massive spin s field
and one massless field with a different spin $s'$. Later, we restrict the general approach to the case of $s'=2$ for Kerr, or to $s'=1$ for root-Kerr in examples.
With this set up, we introduce quartic interactions between the 
fields by adding the quartic vertices to the cubic Lagrangian and by deforming the cubic gauge transformation rules
with the relevant terms.
We present defining equations for the quartic vertices, demonstrate gauge invariance and gauge algebra closure at the quartic level,
and give example solutions for some simple choices of cubic interactions relevant to the Kerr black-hole scattering problem. Since, as expected, these equations
are fairly complicated, we shall also
 consider a simpler problem of on-shell interactions.\newline

The paper is organized as follows.
    Section \ref{sec:brstcubic}
    contains a brief account of 
    the results obtained in \cite{Skvortsov2024}.
We collect some  
 main facts from the gauge invariant BRST formulation
    of massless and massive higher spin fields,
    when  corresponding Lagrangians contain a free part as well as cubic interactions. We also list the BRST invariant cubic vertices corresponding to the on-shell three point functions obtained in \cite{ArkaniHamed2017}.

    In section \ref{sec:aqi} we add quartic interactions
    to the system 
    described in section \ref{sec:brstcubic}.
    We consider three types of quartic vertices: the main vertex of interest to us which includes two copies of the massive field and two copies of the massless field ($W$); and quartic vertices containing four copies of each, the massive field ($U$), and the massless field ($X$), which are required for consistency. We derive the system of equations 
    these vertices should satisfy in order to maintain 
    both the gauge invariance of the Lagrangian
    and formal closure of the gauge algebra. 

    In section \ref{sec:sol} we show that this system of equations for the quartic vertices is consistent and provide a formal particular solution, given that the cubic 
    vertices are BRST invariant; a requirement that is
    satisfied by construction by all consistent cubic 
    interactions. Although this solution is written in 
    an apparent nonlocal form, local solutions do exist as well. We give an explicit local solution for the interaction between a complex massive vector field and a massless vector field.
    
    In section \ref{seconshell} we simplify the off-shell approach by restricting
    the fields to satisfy the free equations of motion. In other words, we consider on-shell quartic interactions. We present the defining equations for on-shell vertices and give an explicit local solution for the case of interaction between a massive vector field and a graviton. The existence of this local solution hinges on adding the graviton cubic vertex with a specific choice of coupling constant.
    
    Finally, some lengthy computations and technical details are collected in the Appendices.

    \section{Cubic Action for Kerr Black Holes }
    \label{sec:brstcubic}

    We start by reviewing the construction of gauge invariant Lagrangians, which describe cubic interactions of three bosonic fields with a priori different masses and spins (see \cite{Fotopoulos:2008ka} for a detailed review). The three bosonic fields 
    present in  cubic interaction vertices
    are represented as ket-vectors $\ket{\phi}$ in an auxiliary Fock-space. We take three copies of this Fock-space, one for each field, labeled by $i=1,2,3$. The Fock-spaces are spanned by commuting creation and annihilation operators
    \begin{equation} \label{eq:brstcubic:OscillatorsBosonic}
        \left[\alpha_{i}^\mu,\alpha^{\dagger\nu}_j\right]=-\eta^{\mu\nu}\delta_{ij}\ ,\quad 
        [ \xi_i, \xi^\dagger_j] = \delta_{ij}\ , \quad \xi_i \ket{0}=0\ , 
    \end{equation}
    \begin{equation}\label{eq:brstcubic:VacChoiceBosonic}
    		 \alpha^\mu_i\ket{0} =\xi_i\ket{0}=0\ ,
    \end{equation}
    with the signature of the $D$ - dimensional Minkowski metric $\eta_{\mu\nu}$ being ``mostly minus". The  ladder operators $\xi_i$, $\xi_i^\dagger$ 
    are needed 
    to describe the ''extra" degree of freedom present in massive fields compared to massless ones. 
    
    A reducible spin-$s$ massive bosonic field can be described as a single Fock-space vector expanded in these oscillators
    \begin{equation} \label{eq:brstcubic:FockSpaceVectorDef}
        \ket{\varphi_i}=\frac{1}{s!}\sum_{k=0}^s\varphi_i^{\mu_1\ldots\mu_{s-k}}(x)\alpha_{i,\mu_1}^\dagger\ldots\alpha_{i,\mu_{s-k}}^\dagger\left(\xi^\dagger_{i}\right)^k\ket{0}\ ,
    \end{equation}
    where $\varphi_i^{\mu_1\ldots\mu_{s-k}}(x)$ are symmetric fields. Massless fields take on the same expansion, but with the simplifying substitution $\xi_i^\dagger\rightarrow0$. We further expand the auxiliary Fock-space by introducing anticommuting ghost-oscillators
    \begin{equation} \label{eq:brstcubic:ghosts}
    	\left\{c^0_i,b^0_j\right\}=\left\{c_i,b_j^\dagger\right\}=\left\{c^\dagger_i,b_j\right\}=\delta_{ij}\ ,
    \end{equation}
    \begin{equation}\label{eq:brstcubic:VacChoiceGhost}
             b^0_i\ket{0}=b_i\ket{0}=c_i\ket{0}=0\ ,
    \end{equation}
    where the $c$-ghosts are assigned ghost number $+1$ and the $b$-ghosts are assigned ghost number $-1$. The field vectors $\ket{\phi_i}$, which are required to have ghost-number $0$, can then be decomposed into
    \begin{equation} \label{eq:brstcubic:triplet}
        \ket{\phi_i}=\ket{\varphi_i}+c^0_ib^\dagger_i\ket{C_i}+c_i^\dagger b^\dagger_i\ket{D_i}\ ,
    \end{equation}
    where the spin-$s$, $s-1$, and $s-2$ component fields $\ket{\varphi_i}$, $\ket{C_i}$, and $\ket{D_i}$ do not contain ghost oscillators. The field $\ket{\varphi_i}$ encodes the actual physical degrees of freedom. It contains a set of irreducible bosonic fields with spins $s, s-2,...,1/0$, depending on whether the highest spin is even or odd. We shall simply refer to $\ket{\varphi_i}$ and $\ket{\phi_i}$ as spin-$s$ fields, according to their highest spin component, leaving the fact that they are reducible implicit. The fields $\ket{C_i}$ and $\ket{D_i}$ are auxiliary and do not carry any physical degrees of freedom (see \cite{Skvortsov2024, Pashnev:1997rm} for a detailed discussion).
    
    We are now ready to combine these gauge fields into a cubic Lagrangian which takes the form
    \begin{equation} \label{cubicL1}
    		\mathcal{L}=-\sum_{i=1}^{3}\int dc^0_i\braket{\phi_i|Q_i|\phi_i}+g\left(\int dc^0_1dc^0_2dc^0_3\bra{\phi_1}\bra{\phi_2}\bra{\phi_3}\ket{V}+c.c.\right)\ .
    \end{equation}
    In this equation $Q_i$ are BRST charges, $\ket{V}$ is a cubic vertex and $g$ is a coupling constant.
     The BRST charges 
     \begin{equation} \label{eq:brstcubic:BRSTCharge}
    		Q_i=c^0_il^0_i+c_i\,l_i^\dagger+c_i^\dagger l_i+b^0_ic_ic^\dagger_i\ ,
    \end{equation}
     are functions of the bosonic oscillators and anti-hermitian momenta $p_{i,\mu} \equiv \partial_{i,\mu}$, using\footnote{We are using shorthand $a\cdot b \equiv a_\mu b^\mu$.}
    \begin{equation}
        l^0_i\equiv p_i\cdot p_i+m_i^2\ ,\quad l_i\equiv p_i\cdot\alpha_i-m_i\xi_i\ ,\quad l_i^\dagger\equiv- p_i\cdot\alpha_i^\dagger-m_i\xi^\dagger_i\ ;
    \end{equation}
    with no summation over repeated indices $i$. At the linearized level, this defines physical states as BRST closed vectors, $Q_i\ket{\phi_i}=0$. Our choice of vacuum~\eqref{eq:brstcubic:VacChoiceGhost} ensures that this condition enforces transversality $l_i\ket{\varphi_i}=0$ and the mass-shell condition $l^0_i\ket{\varphi}=0$ on the physical fields.
    The Lagrangian~\eqref{cubicL1} is required to be invariant under gauge transformations
    \begin{equation} \label{gtcubic}
        \delta\ket{\phi_i}=Q_i\ket{\Lambda_i}+g\int dc^0_{i+1}dc^0_{i+2}\left(\bra{\phi_{i+1}}\bra{\Lambda_{i+2}}+\bra{\Lambda_{i+1}}\bra{\phi_{i+2}}\right)\ket{V}\ ,
    \end{equation}
    up to the first order in the coupling constant $g$. The gauge parameters $\ket{\Lambda_i}=b_i^\dagger\ket{\lambda}$ with $\ket{\lambda}$ contain only
    $\alpha_{i,\mu}^\dagger$ and $\xi_{i}^\dagger$ oscillators of total degree
    ($s$-1)  (for a spin-$s$ $\ket{\phi_i}$). The invariance of~\eqref{cubicL1} as well as the closure of the gauge algebra up to first order in $g$ is enforced by the nilpotency of the BRST charges $Q_i^2=0$ and by a single condition on the cubic vertex $\ket{V}$
    \begin{equation}\label{eq:brstcubic:BRSTInvarianceCubic}
    	\left(Q_1+Q_2+Q_3\right)\ket{V}=0\ .
    \end{equation}
    
    Now let us turn to the particular case of cubic interactions between a field with spin $s$ and mass $m$ and a massless spin $s'$ field. We will put the massive field in Fock-spaces $i=1,2$ and require that the component fields of $\ket{\phi_1}$ and $\ket{\phi_2}$ are complex conjugates of each other (e.g., $\varphi_{1,\mu_1\ldots\mu_s}(x)=\varphi_{2,\mu_1\ldots\mu_s}^\star(x)$). When the massless field has spin $2$, we further require those component fields to be real. The massless field is placed in Fock-space $i=3$.
    
    A general solution for a vertex $\ket{V}$ which satisfies the equation~\eqref{eq:brstcubic:BRSTInvarianceCubic} was given in \cite{Metsaev:2012uy}
    \begin{equation}
        \ket{V}=V(\mathcal{K}_i,\mathcal{Q},\mathcal{Z})c_1^0c_2^0c_3^0\ket{0}\ ,
    \end{equation}
    where $V(\mathcal{K}_i,\mathcal{Q},\mathcal{Z})$ is a function of the five BRST invariants
    \begin{align}
        \mathcal{K}_1&=-(p_2-p_3)\cdot\alpha_1^\dagger+m\xi_1^\dagger+(b_2^0-b_3^0)c_1^\dagger\ ,\label{eq:brstcubic:K1}\\
        \mathcal{K}_2&=-(p_3-p_1)\cdot\alpha_2^\dagger-m\xi_2^\dagger+(b_3^0-b_1^0)c_2^\dagger\ ,\\
        \mathcal{K}_3&=-(p_1-p_2)\cdot\alpha_3^\dagger+(b_1^0-b_2^0)c_3^\dagger\ ,\\
        \mathcal{Q}&=\mathcal{Q}^{(1,2)}+\frac{\xi_1^\dagger}{2m}\mathcal{K}_2-\frac{\xi_2^\dagger}{2m}\mathcal{K}_1\ ,\\
        \mathcal{Z}&=\mathcal{Q}^{(1,2)}\mathcal{K}_3+\mathcal{Q}^{(2,3)}\mathcal{K}_1+\mathcal{Q}^{(3,1)}\mathcal{K}_2\ , \label{eq:brstcubic:cZ}
    \end{align}
    with 
	\begin{align}
		\mathcal{Q}^{(1,2)}&=-\alpha_1^\dagger\cdot\alpha_2^\dagger+\xi_1^\dagger\xi_2^\dagger-\frac{1}{2}\left(b_1^\dagger c_2^\dagger + b_2^\dagger c_1^\dagger\right)\ ,\\
		\mathcal{Q}^{(i,i+1)}&=-\alpha_i^\dagger\cdot\alpha_{i+1}^\dagger-\frac{1}{2}\left(b_i^\dagger c_{i+1}^\dagger + b_{i+1}^\dagger c_i^\dagger\right)\ ,\quad i=2,3\ .
	\end{align}
    We are specifically interested in cubic vertices which reproduce the  amplitudes
    describing minimal coupling
    ~\cite{ArkaniHamed2017} which have been linked to Kerr-black hole scattering. As was shown in \cite{Skvortsov2024}, in order to reproduce the $s-s-2$ minimal coupling amplitude, one has to choose $V$ as
    \begin{equation} \label{cubKerr}
        V= ({\cal K}^{(3)})^2+
        \left ( {\cal Z}{\cal K}^{(3)}+  \frac{{\cal Z}^2 -
        {\cal Q}^2 {\cal K}^{(3)} {\cal Z}}{ (1 - {\cal Q})^2  - \frac{1}{2m^2} 
        {\cal K}^{(1)} {\cal K}^{(2)} }
        \right)\ .
    \end{equation}
    For the root-Kerr problem which corresponds to the case $s-s-1$, i.e. massive spin $s$ interacting with a massless spin $1$ field, minimal coupling is obtained by the choice
    \be \label{cubrKerr}
        V=i{\cal K}^{(3)}+i\frac{{\cal Z} -
        {\cal Q}^2 {\cal K}^{(3)}
        }{ (1 - {\cal Q})^2  - \frac{1}{2m^2} 
        {\cal K}^{(1)} {\cal K}^{(2)}}\ .
    \ee
    The expressions \p{cubKerr} and \p{cubrKerr}
    are actually off-shell extensions
    of the on-shell generating functionals for three point interactions obtained in \cite{Chiodaroli:2021eug} written in the four dimensional spinor helicity formalism.

    For the purpose of extending the above results to quartic interactions, we also need to introduce cubic interactions between three massless fields. To this end, one essentially repeats the entire construction
    given above with all three fields $\ket{\phi_i}$ chosen to be massless (i.e., discard dependence on $\xi$ and $m$ everywhere). We call this massless cubic vertex $\Vgr$ since the case where the massless field is spin 2 will be of specific interest to us. Any BRST invariant $\Vgr$ is admissible in our description, but for examples we will focus on the case where it describes the three graviton vertex 
    \be \label{eq:brstcubic:3gr}
        \Vgr =  g_2 {\tilde{\mathcal{Z}}}^2\ ,
    \ee
    where $\tilde{\mathcal{Z}}$ is obtained from \eqref{eq:brstcubic:cZ} by discarding all $\xi$ and $m$ dependence.
    A priori, the coupling constant $g_2$ in \eqref{eq:brstcubic:3gr} is independent from the coupling constant $g$ in \eqref{cubicL1}. As we shall see in section~\ref{sec:112}, the requirement of locality of quartic interactions imposes a relationship between these coupling constants.
    
    Finally, we can use the fact that the vertices \eqref{cubKerr}, \eqref{cubrKerr}, and \eqref{eq:brstcubic:3gr} satisfy a generic reality condition. We take \eqref{cubrKerr} as an example, and make the dependence on the Fock-space oscillators and momenta explicit $V(\bar{1},\bar{2},\bar{3},p_1,p_2,p_3)$, i.e. $V$ depends on creation operators of spaces $1$, $2$, and $3$, as well as corresponding momenta. Then the vertex \eqref{cubrKerr} satisfies the reality condition (see Appendix
    \ref{sec:bravsket}
    for details)
    \begin{equation}\label{eq:brstcubic:cubicRealCond}
        V(\bar{1},\bar{2},\bar{3},p_1,p_2,p_3) = V^\dagger(\bar{2},\bar{1},\bar{3},-p_2,-p_1,-p_3)\vert_{c_i^\dagger\rightarrow -c_i^\dagger}\ ,
    \end{equation}
    This equation means that we first take the hermitian conjugate of the vertex, then replace all annihilation operators with creation ones (with the exception of $b_i^0$ which does not get replaced with $c_i^0$), swap all oscillators and momenta in the massive oscillator space, then add a minus sign to all momenta and $c_i^\dagger$. The invariance under this operation guarantees the reality of the Lagrangian and of the coupling constants 
    when $\ket{\phi_i}$ contain
     real fields.
     This condition is satisfied by all three vertices \eqref{cubKerr}, \eqref{cubrKerr}, and \eqref{eq:brstcubic:3gr}. The condition~\eqref{eq:brstcubic:cubicRealCond} implies that the cubic term in~\eqref{cubicL1} is already real, and we can write the Lagrangian in the form
    \begin{equation} \label{cubicL2}
        \mathcal{L}=-\sum_{i=1}^{3}\int dc^0_i\braket{\phi_i|Q_i|\phi_i}+2g\int dc^0_1dc^0_2dc^0_3\bra{\phi_1}\bra{\phi_2}\bra{\phi_3}\ket{V}\ .
    \end{equation}
 which is more convenient for subsequent
 computations.
    
    \section{Adding Quartic Interactions} \label{sec:aqi}
    
    Let us proceed by describing the quartic extension\footnote{See 
    \cite{Fotopoulos:2010ay, Taronna:2011kt, Dempster:2012vw}
     for a construction of quartic interactions for massless fields
    in the BRST approach  
    and \cite{Bengtsson:2006pw} for a
     general procedure for determining higher order interactions for massless higher spin fields 
     in the BRST approach.} 
    of the Lagrangian
    \p{cubicL2}. This procedure contains  
    several
    steps. First, we add quartic vertices to the cubic Lagrangian, then we describe how to extend the gauge transformations in a  manner that is consistent with the added quartic interactions. Next, we derive conditions for the quartic vertices, similar to~\eqref{eq:brstcubic:BRSTInvarianceCubic}, that ensure both invariance of the Lagrangian and gauge algebra closure. 
    
    For the purpose of this quartic extension the treatment of the massive field in Fock-spaces $i=1$, and $i=2$ is unchanged from section \ref{sec:brstcubic}. Regarding the massless field, we add some more copies.
    We place the massless field in Fock-spaces $i=3$, $4$, and $5$ and require the component fields of those three Fock-spaces to be the same (e.g., $\varphi_3^{\mu\nu}(x)=\varphi_4^{\mu\nu}(x)=\varphi_5^{\mu\nu}(x)$). As we mentioned in the previous section, in general we also include the massless cubic vertex \p{eq:brstcubic:3gr} (note that e.g. for root-Kerr, i.e. spin 1 massless field, this vertex is vanishing). Let us stress that in what follows we never use the explicit form of the cubic vertex given in equations \p{cubKerr}-- \p{cubrKerr}. Rather, we assumed only the symmetry properties of these vertices and the reality condition~\eqref{eq:brstcubic:cubicRealCond}.

    When varying the Lagrangian \p{cubicL2} with respect to gauge transformations~\p{gtcubic} one obtains at order $g^2$ terms of the form
	\begin{equation}
		\begin{split}
		\delta\mathcal{L}=&\ g(\ldots)-2g^2\sum_{i=1}^{3}\int dc^0_1dc^0_2dc^0_3\times\\
		&\ \bra{V}\ket{\phi_{i+1}}\ket{\phi_{i+2}}\int dc^0_{i'+1}dc^0_{i'+2}\left(\bra{\Lambda_{i'+1}}\bra{\phi_{i'+2}}+\bra{\phi_{i'+1}}\bra{\Lambda_{i'+2}}\right)\ket{V}\ .
		\end{split}
	\end{equation}
	As described above, we expand the Lagrangian with additional copies of the massless fields and include quartic vertices which can compensate these terms at order $g^2$
    \begin{equation}\label{LQ}
		\begin{split}
		\mathcal{L}=-\sum_{i=1}^{4}\int dc^0_i\braket{\phi_i|Q_i|\phi_i}&+4g\int dc^0_1dc^0_2dc^0_3\bra{\phi_1}\bra{\phi_2}\bra{\phi_3}\ket{V}\\
        &+\frac{2}{3}g_2\int dc_3^0dc_4^0dc_5^0\bra{\phi_3}\bra{\phi_4}\bra{\phi_5}\ket{\Vgr}\\
		&+2g^2\int dc^0_1dc^0_2dc^0_3dc^0_4\bra{\phi_1}\bra{\phi_2}\bra{\phi_3}\bra{\phi_4}\ket{W}\\
		&+g^2\int dc^0_1dc^0_2dc^0_{1'}dc^0_{2'}\bra{\phi_1}\bra{\phi_2}\bra{\phi_{1'}}\bra{\phi_{2'}}\ket{U}\\
        &+g_2^2\int dc_3^0dc^0_4dc_{3'}^0dc_{4'}^0\bra{\phi_3}\bra{\phi_4}\bra{\phi_{3'}}\bra{\phi_{4'}}\ket{X}\ .
		\end{split}
	\end{equation}
As can be seen from the Lagrangian \p{LQ},
there are three types of quartic vertices:
\begin{itemize}
    \item The vertex $\ket{W}$
    describes interactions between two massive and two massless fields.
\item The vertex $\ket{U}$
    describes interactions between four massive  fields.

   \item The vertex $\ket{X}$
    describes interactions between four  massless fields.
    
\end{itemize}
    All quartic vertices have ghost-number $4$ and have the form
    \be
    \ket{W}\equiv Wc_1^0c_2^0c_{3}^0c_{4}^0\ket{0}\ ;\quad\ket{U}\equiv Uc_1^0c_2^0c_{1'}^0c_{2'}^0\ket{0}\ ;\quad\ket{X}\equiv Xc_3^0c_4^0c_{3'}^0c_{4'}^0\ket{0}\ ,
    \ee
    where $W,\ U,$ and $X$ are ghost-number zero functions of the momenta $p_{i,\mu}$, bosonic creation operators $\alpha_{i,\mu}^\dagger$, $\xi^\dagger_i$, ghost creation operators $c_i^\dagger$, $b_i^\dagger$ and annihilation operators $b_i^0$. The four vertices are required to satisfy reality conditions analogous to~\eqref{eq:brstcubic:cubicRealCond}
    \begin{align}
        \label{eq:aqi:RealCond}
        W(\bar{1},\bar{2},\bar{3},\bar{4},p_1,p_2,p_3,p_4) &= W^\dagger(\bar{2},\bar{1},\bar{3},\bar{4},-p_2,-p_1,-p_3,-p_4)\vert_{c_i^\dagger\rightarrow -c_i^\dagger}\ ,\\
        U(\bar{1},\bar{2},\bar{1}',\bar{2}',p_1,p_2,p_1',p_2') &= U^\dagger(\bar{2},\bar{1},\bar{2}',\bar{1}',-p_2,-p_1,-p_2',-p_1')\vert_{c_i^\dagger\rightarrow -c_i^\dagger}\ ,\\
        X(\bar{3},\bar{4},\bar{3}',\bar{4}',p_3,p_4,p_3',p_4') &= X^\dagger(\bar{3},\bar{4},\bar{3}',\bar{4}',-p_3,-p_4,-p_3',-p_4')\vert_{c_i^\dagger\rightarrow -c_i^\dagger}\ .
    \end{align}
    Next, we require the Lagrangian \p{LQ} to be invariant under 
    the gauge transformations 
	\begin{align}\label{eq:offshell_gaugetransf1}
		\delta\ket{\phi_1}=Q_1\ket{\Lambda_1}+2g&\int dc^0_{2}dc^0_{3}\left(\bra{\phi_{2}}\bra{\Lambda_{3}}+\bra{\Lambda_{2}}\bra{\phi_{3}}\right)\ket{V}+g^2\delta_U\ket{\phi_1}+g^2\delta_W\ket{\phi_1}\\
        \label{eq:offshell_gaugetransf2}%
		\delta\ket{\phi_2}=Q_2\ket{\Lambda_2}+2g&\int dc^0_{3}dc^0_{1}\left(\bra{\phi_{1}}\bra{\Lambda_{3}}+\bra{\Lambda_{1}}\bra{\phi_{3}}\right)\ket{V}+g^2\delta_U\ket{\phi_2}+g^2\delta_W\ket{\phi_2}\\
        \begin{split}\label{eq:offshell_gaugetransf3}
			\delta\ket{\phi_3}=Q_3\ket{\Lambda_3}+\ g&\int dc^0_{1}dc^0_{2}\left(\bra{\phi_{1}}\bra{\Lambda_{2}}+\bra{\Lambda_{1}}\bra{\phi_{2}}\right)\ket{V}+g_2^2\delta_X\ket{\phi_3}+g^2\delta_W\ket{\phi_3}\\
            +\;g_2&\int dc^0_{4}dc^0_{5}\bra{\Lambda_{4}}\bra{\phi_{5}}\ket{\Vgr}
        \end{split}
	\end{align}
     The variation of $\ket{\phi_4}$, and $\ket{\phi_5}$ is identical to that of $\ket{\phi_3}$ with  $3\leftrightarrow 4$ and  $3\leftrightarrow 5$,  respectively. The additional part of the variation which is proportional to $g^2$ is given by
    \begin{equation}\label{eq:offshell_gaugetransf}
    \begin{split}
        \delta_F\ket{\phi_i}=(-1)^i\int &dc^0_{i+1}dc^0_{i+2}dc^0_{i+3}\Big(\bra{\Lambda_{i+1}}\bra{\phi_{i+2}}\bra{\phi_{i+3}}+\\ 
        &+\bra{\phi_{i+1}}\bra{\Lambda_{i+2}}\bra{\phi_{i+3}}+\bra{\phi_{i+1}}\bra{\phi_{i+2}}\bra{\Lambda_{i+3}}\Big)\ket{F}\ ,
    \end{split}
	\end{equation}
    with $(i,i+1,i+2,i+3)$ identified with the cyclic permutation $(1,2,3,4)$ when $F=W$, $(1,2,1',2')$ when $F=U$, or $(3,4,3',4')$ when $F=X$ respectively.

    The requirement of the Lagrangian~\eqref{LQ} to be invariant under gauge transformations \eqref{eq:offshell_gaugetransf1}-\eqref{eq:offshell_gaugetransf3} imposes conditions on the quartic vertices. Contrary to the condition arising at  cubic level \eqref{eq:brstcubic:BRSTInvarianceCubic}, this results in inhomogeneous equations. These inhomogeneities are quadratic in the cubic vertices and are formed by a ''contraction" of one of the Fock-spaces in each vertex. 
    
    Before writing down the equations, it will be useful to 
introduce short hand notations for 
 these contractions. We define for any cubic vertex $V$
	\begin{align}
		\begin{split}\label{eq:dW:ContrDef1}
		\ket{V:_1V}&\equiv\int dc_{1'}^0\bra{0}_{1'} c_{1'}^0\left( V(\bar{1},1',\bar{4},p_1,-p_1-p_4,p_4)\big\vert_{c_{1'}\rightarrow-c_{1'}}\right)\times\\
		&\hspace{3.3cm}V(\bar{1}',\bar{2},\bar{3},-p_2-p_3,p_2,p_3)c_{1'}^0c_1^0c_2^0c_3^0c_4^0\ket{0}\ ;
		\end{split}\\
		\begin{split}\label{eq:dW:ContrDef2}
		\ket{V:_2V}&\equiv\int dc_{2'}^0\bra{0}_{2'} c_{2'}^0\left( V(2',\bar{2},\bar{4},-p_2-p_4,p_2,p_4)\big\vert_{c_{2'}\rightarrow-c_{2'}}\right)\times\\
		&\hspace{3.3cm}V(\bar{1},\bar{2}',\bar{3},p_1,-p_1-p_3,p_3)c_{1}^0c_{2'}^0c_2^0c_3^0c_4^0\ket{0}\ ;
		\end{split}\\
		\begin{split}\label{eq:dW:ContrDef3}
		\ket{V:_3V}&\equiv\int dc_{5}^0\;\bra{0}_{5}\; c_{5}^0\;\left( V(\bar{1},\bar{2},\;5,p_1,p_2,-p_1-p_2)\big\vert_{c_{5}\;\rightarrow-c_{5}}\;\right)\times\\
		&\hspace{3.3cm}\tilde{V}(\bar{1}',\bar{2}',\bar{5},p_{1'},p_{2'},-p_{1'}-p_{2'})c_{1'}^0c_{2'}^0c_1^0c_2^0c_5^0\ket{0}\ .
		\end{split}
	\end{align}
 Whenever we need to compute the expression $\ket{V:_3\Vgr}$,
    we
     simply replace the second  $V$ in the equations above with $\Vgr$ and change $1'\leftrightarrow3$, $2'\leftrightarrow4$.
     The case $\ket{\Vgr:_3V}$ is defined analogously, but replacing the first $V$ and Fock-spaces $1$ and $2$ instead. 
     
     These contractions obey various useful symmetry properties. Using the symmetry of the cubic vertices $V(\bar{1},\bar{2},\bar{3})=(\pm 1)V(\bar{2},\bar{1},\bar{3})$ as well as the technique described in  Appendix~\ref{sec:bravsket}, one finds 
    \begin{equation}\label{eq:dW:contractionSymmetry}
    \begin{split}
        \ket{V:_1V}\big\vert_{\begin{bsmallmatrix}1&\leftrightarrow&2\end{bsmallmatrix}}&=\ket{V:_2V}\ ,\quad%
        \ket{V:_1V}\big\vert_{\begin{bsmallmatrix}3&\leftrightarrow&4\end{bsmallmatrix}}=\ket{V:_2V}\ ,\\
        \ket{V:_3V}\vert_{\begin{bsmallmatrix}1&\leftrightarrow&1'\\
        2&\leftrightarrow&2'\end{bsmallmatrix}}&=\ket{V:_3V}\ ,\quad%
        \ket{V:_3\Vgr}=\ket{\Vgr:_3V}\ .
    \end{split}
    \end{equation}
    Now we are ready to present conditions for the quartic vertices. Using the above contractions and their symmetries, one obtains that the quartic vertices have to satisfy the equations
    \begin{align}\label{eq:dW:Weq}
        \left(Q_1+Q_2+Q_3+Q_4\right)\ket{W}&=-2\ket{V:_1V}+2\ket{V:_2V}-\frac{g_2}{g}\ket{V:_3\Vgr}\ ,\\ \label{eq:dW:Ueq}
        \left(Q_1+Q_2+Q_{1'}+Q_{2'}\right)\ket{U}&=-\left(\ket{V:_3V} -(1\leftrightarrow1')\right)\ ,\\ \label{eq:dW:Xeq}
        \left(Q_3+Q_4+Q_{3'}+Q_{4'}\right)\ket{X}&=-\frac{1}{4}\left(\ket{\Vgr:_3\Vgr}-(3\leftrightarrow3')\right)\ .
    \end{align}
    As in the case 
    of the cubic Lagrangians
    reviewed in  section \ref{sec:brstcubic},
     these equations guarantee that the Lagrangian~\eqref{LQ} is invariant under the gauge transformations \p{eq:offshell_gaugetransf1}--\p{eq:offshell_gaugetransf3} as well as ensuring that the algebra of gauge transformations is associative at order $g^2$.
     We refer to  Appendix~\ref{sec:QuarticVariation} for details of the proof at both cubic and quartic orders. 
    
    \section{Off-shell Solutions for Quartic Vertices}\label{sec:sol}
    \subsection{Formal Particular Solution}

    Below we shall construct formal particular solutions for the quartic vertices $W$, $U$, and $X$,  following the approach in ~\cite{Taronna:2011kt} for massless fields. We do this in two steps. 
    First, we show that the right-hand side (RHS) of the equations~\eqref{eq:dW:Weq}---\eqref{eq:dW:Xeq}
    are BRST closed, which is a necessary condition for the solution to exist. Second, we use this property to construct a solution for the quartic vertex.

    We demonstrate that the RHS of equations \eqref{eq:dW:Weq}---\eqref{eq:dW:Xeq} are BRST closed, by showing that the individual contractions of cubic vertices are BRST closed
    \begin{align}
            (Q_1+Q_2+Q_3+Q_4)\ket{V:_1V}&=0\ ;\label{eq:consistency:cont1}\\
            (Q_1+Q_2+Q_3+Q_4)\ket{V:_2V}&=0\ ;\label{eq:consistency:cont2}\\
            (Q_1+Q_2+Q_3+Q_4)\ket{V:_3V}&=0\ \label{eq:consistency:cont3};\\
            (Q_1+Q_2+Q_3+Q_4)\ket{V:_3\Vgr}&=0\ .\label{eq:consistency:cont4}
    \end{align}
    Because of symmetries~\eqref{eq:dW:contractionSymmetry} the condition~\eqref{eq:consistency:cont2} follows directly from~\eqref{eq:consistency:cont1}. Let us show the latter. From the BRST  invariance of the cubic vertex~\eqref{eq:brstcubic:BRSTInvarianceCubic} we find
    \begin{align}
        &\left[Q_{1'}\vert_{p_{1'}\rightarrow -p_2-p_3}+Q_2+Q_3,V(\bar{1}',\bar{2},\bar{3},-p_2-p_3,p_2,p_3)\right]=0\ ,\\
        &\left[Q_{1'}\vert_{p_{1'}\rightarrow p_1+p_4}+Q_1+Q_4,V(\bar{1},1',\bar{4},p_1,-p_1-p_4,p_4)\big\vert_{c_{1'}\rightarrow-c_{1'}}\right]=0\ .
    \end{align}
    The first of these follows directly. For the second one, we use the techniques of Appendix \ref{sec:bravsket} since $V(1',\bar{1},\bar{4})$ has annihilation operators instead of creation operators in its first oscillator space. Both are also easily verified for individual invariants~\eqref{eq:brstcubic:K1}-\eqref{eq:brstcubic:cZ}. With this we can straightforwardly compute 
    \begin{equation}
    \begin{split}
        &(Q_2+Q_3)\ket{V:_1V}=\\
        &\int dc_{1'}^0\bra{0}_{1'} c_{1'}^0\left( V(\bar{1},1',\bar{4},p_1,-p_1-p_4,p_4)\vert_{c_{1'}\rightarrow-c_{1'}}\right)\times\\
		&\hspace{3cm}(-Q_{1'}\vert_{p_{1'}\rightarrow -p_2-p_3})V(\bar{1}',\bar{2},\bar{3},-p_2-p_3,p_2,p_3)c_{1'}^0c_1^0c_2^0c_3^0c_4^0\ket{0}=\\
        &\int dc_{1'}^0\bra{0}_{1'} c_{1'}^0\left( V(\bar{1},1',\bar{4},p_1,-p_1-p_4,p_4)\vert_{c_{1'}\rightarrow-c_{1'}}\right)\times\\
		&\hspace{3cm}(-Q_{1'}\vert_{p_{1'}\rightarrow p_1+p_4})V(\bar{1}',\bar{2},\bar{3},-p_2-p_3,p_2,p_3)c_{1'}^0c_1^0c_2^0c_3^0c_4^0\ket{0}=\\
        &-(Q_1+Q_4)\ket{V:_1V}\ ,
    \end{split}
    \end{equation}
    where we used momentum conservation in the second equality. We have therefore shown~\eqref{eq:consistency:cont1}. The proof for~\eqref{eq:consistency:cont3} and~\eqref{eq:consistency:cont4} is completely identical.

    We are now ready to construct a solution to~\eqref{eq:dW:Weq}. Making the observation that
    \begin{equation}
        l_i^0=\left\{Q_i,b_i^0\right\}\ .
    \end{equation}
    It therefore follows directly that for $\ket{RHS}\equiv-2\ket{V:_1V}+2\ket{V:_2V}-\frac{g_2}{g}\ket{V:_3\Vgr}$, we have
    \begin{equation}
        \begin{split}
            \ket{RHS}&=\frac{1}{l_4^0}\left\{Q_4,b_4^0\right\}\ket{RHS}\\
            &=\frac{1}{l_4^0}\left(Q_4b_4^0-b_4^0(Q_1+Q_2+Q_3)\right)\ket{RHS}\\
            &=\left(Q_1+Q_2+Q_3+Q_4\right)\frac{1}{l_4^0}b_4^0\ket{RHS}\ ,
        \end{split}
    \end{equation}
    which respecting the symmetries of $W$ and~\eqref{eq:dW:Weq} gives us the result
    \begin{equation}
        \ket{W}=\frac{1}{2}\left(\frac{1}{l_3^0}b_3^0+\frac{1}{l_4^0}b_4^0\right)\ket{RHS}\ .
    \end{equation}
    The exact same type of argument also gives 
    \begin{align}
        \ket{U}&=-\frac{1}{4}\left(\frac{1}{l_1^0}b_1^0+\frac{1}{l_2^0}b_2^0+\frac{1}{l_{1'}^0}b_{1'}^0+\frac{1}{l_{2'}^0}b_{2'}^0\right)\Big(\ket{V:_3V}-\left(1\leftrightarrow1'\right)\Big)\ ;\\
        \ket{X}&=-\frac{1}{16}\left(\frac{1}{l_3^0}b_3^0+\frac{1}{l_4^0}b_4^0+\frac{1}{l_{3'}^0}b_{3'}^0+\frac{1}{l_{4'}^0}b_{4'}^0\right)\Big(\ket{\Vgr:_3\Vgr}-\left(3\leftrightarrow3'\right)\Big)\ .
    \end{align}
    Apparently, this solution is not unique, since one can always add 
    a solution of the homogeneous equation to it (i.e., of the equations
    \p{eq:dW:Weq}--\p{eq:dW:Xeq}
    with the right hand sides set equal to zero). Furthermore, the above solution is written in an apparently non-local form. However, as  shown in the examples below, manifestly local solutions for quartic vertices do exist as well. The existence of such local solutions appears to depend concretely on the form of the three-point vertices. In fact, as will be shown explicitly in section~\ref{sec:112}, when the cubic interaction is described by a vertex which contains two massive vector fields and one massless spin $2$ field (i.e., the $1-1-2$ minimal coupling vertex) the existence of a local quartic vertex requires the addition of the graviton cubic vertex with a ``fine tuned" value of the coupling constant. Finally, let us note that a solution
    of the homogeneous equations can also contribute
    to cancellation of nonlocalities (see \cite{Taronna:2011kt} for a similar discussion for the case of massless fields). This fact is explicit in the next section, where we present a local solution for the spin $1-1-1$ minimal coupling vertex.
    
    \subsection{Example: Quartic Solution for 1-1-1 Cubic Vertex}\label{sec:111}
    
    We present a local solution for quartic vertices here. The procedure for computing them is similar to that for cubic vertices \cite{Buchbinder:2006eq} and can be summarized as follows.
    
    The functions $W$, $U$ and $X$ are Lorentz scalars and have ghost number zero. Therefore, they can be some polynomial in terms of the bilinears
    \be
    p_i \cdot p_j, \quad\alpha^\dagger_i \cdot \alpha^\dagger_j,
    \quad 
    p_i \cdot \alpha^\dagger_j,
    \quad 
    c^\dagger_i b_j^0, \quad c^\dagger_i b^\dagger_j
    \ee
    as well of the operators $\xi_i^\dagger$. The total number of ``daggered" operators in each term of the polynomial is equal to the total spin entering the quartic vertex. The constant coefficients of each terms are to be determined from the defining equations. We note that the number of terms in these expressions grows very rapidly with increasing spin, due to the growing number of oscillators. We can slightly improve on general polynomials of the above oscillator combinations using the symmetries of the vertex and comparing with existing combinations in the cubic vertices (e.g., this can give us an upper limit on the power of momenta appearing in the quartic vertex). The computations in this paper were performed by hand and later confirmed numerically (using symbolic calculations for oscillators and randomized numerical values for momenta to increase speed). In principle, it should be straightforward to apply numerical methods to derive solutions with higher spin than the ones we present here.

    We can apply the above to quartic vertices 
    \p{eq:dW:Weq}--\p{eq:dW:Xeq}
    for the simplest non-trivial example, the spin 1 part of the minimal coupling root-Kerr vertex, i.e. one massive complex vector field interacting with a massless vector field via the cubic vertex
    \begin{equation} \label{Z111}
        \ket{V}=i\mathcal{Z}c_1^0c_2^0c_3^0\ket{0}\ .
    \end{equation}
    We make the following two definitions to increase the readability of explicit expressions in this and the following sections,
    \begin{equation}
        \A{i}{j}\equiv\alpha_{i}^\dagger\cdot\alpha^\dagger_{j}\ ;\quad \Pa{i}{j}\equiv p_{i}\cdot\alpha_{j}^\dagger\ .
    \end{equation}
    Since we have a spin 1 massless field $\Vgr$ and $X$ vanish. The $W$ and $U$ vertices below were computed by solving their respective equations. The quartic interactions that extend the cubic vertex~\eqref{Z111} are
    \begin{equation}\label{W111}
    \begin{split}
        W=2&\A{1}{3}\A{2}{4}+2\A{1}{4}\A{2}{3}-4(\A{1}{2}-\xi_1^\dagger\xi_2^\dagger)\A{3}{4}\\
        +&\frac{1}{2m}\left(\Pa{1}{1}\xi^\dagger_2+\Pa{2}{2}\xi^\dagger_1+2m\xi^\dagger_1\xi^\dagger_2+\xi^\dagger_1b_2^0c_2^\dagger+\xi^\dagger_2b_1^0c_1^\dagger\right)\left(b_3^\dagger c_4^\dagger+b_4^\dagger c_3^\dagger\right)\\
        +&\frac{1}{2m}\left(\left(\Pa{4}{4}+b_4^0c_4^\dagger\right)c_3^\dagger+\left(\Pa{3}{3}+b_3^0c_3^\dagger\right)c_4^\dagger\right)\left(\xi_2^\dagger b_1^\dagger +\xi_1^\dagger b_2^\dagger \right)\ ,
    \end{split}
    \end{equation}
    and
    \begin{equation} \label{U111}
    \begin{split}
        U= \bigg(\bigg(-\frac{1}{8} \Big(&-2\A{1}{1'}\A{2}{2'} + 2\A {1}{2}\A{1'}{2'}-4\A{2}{2'}\xi^\dagger_1\xi^\dagger_{1'} + 4\A {1'}{2}\xi^\dagger_ 1\xi^\dagger_{2'}\\
        &+\frac{1}{m^2}\Pa {2}{2}\Pa {1'}{1'}\xi^\dagger_1\xi^\dagger_{2'}+\frac{8}{m} \Pa {1}{1'}\xi^\dagger_ 1\xi^\dagger_ 2\xi^\dagger_{2'} +\frac{2}{m} \Pa {1'}{1'}\xi^\dagger_ 1\xi^\dagger_ 2\xi^\dagger_{2'}\\
        &+\frac{1}{m^2}(4 p_{1'} \cdot p_{1'} +  4 p_{1'} \cdot p_{2} + m^2)\xi^\dagger_ 1\xi^\dagger_ 2\xi^\dagger_{1'}\xi^\dagger_{2'}\\
        &+\frac{1}{m^2}\left(2\Pa {1}{1}\xi^\dagger_ {1'}\xi^\dagger_{2} + 2m \xi^\dagger_ 1\xi^\dagger_ 2\xi^\dagger_ {1'} + \xi^\dagger_ 1\xi^\dagger_ 2 b^0_{1'}c^\dagger_{1'}\right)b^0_{2'}c^\dagger_{2'}\Big)\\
        +\frac{1}{m}\Big(&-\xi^\dagger_ 1\xi^\dagger_ {1'}\xi^\dagger_{2'} b^0_{1}+\frac{1}{4} \left(\Pa {2'}{2'}\xi^\dagger_ {1'} - 3 m\xi^\dagger_ {1'}\xi^\dagger_ {2'} + \xi^\dagger_ {1'} b^0_{2'}c^\dagger_{2'}\right) b^\dagger_{1}\Big)c^\dagger_{2}\\
      +(1\leftrightarrow&1')\bigg)+(2\leftrightarrow2')\bigg)+(1\leftrightarrow2, 1'\leftrightarrow2')\ .
    \end{split}
    \end{equation}
    The set of vertices \p{W111},\p{U111} along with the free Lagrangian given by the first term in \p{LQ} gives a gauge invariant description of  a complex massive vector field interacting with a massless vector field. This Lagrangian contains cubic and quartic interactions, both of which are local. As a result of a gauge invariant description, one has an auxiliary Stueckelberg massive complex scalar field.

    \section{On-shell Vertices} \label{seconshell}

    \subsection{Defining Equations}
    
    Up to now, we were dealing with off-shell interactions. The equations on quartic vertices in this approach are quite complicated, due to the presence
    of auxiliary fields and due to the unconstrained form of the physical fields. Therefore, it can be helpful to consider on-shell vertices, similarly
    to how it was done in~\cite{Sagnotti:2010at,Dempster:2012vw,Karapetyan:2021wdc}. 
    
    The off-shell equations~\eqref{eq:dW:Weq}--~\eqref{eq:dW:Xeq} follow from the requirement of the Lagrangian~\eqref{LQ} to be gauge invariant. We can obtain their on-shell equivalent by taking the fields appearing in this gauge invariance condition (e.g.,~\eqref{eq:QuarticVariation:Weq}, \eqref{eq:QuarticVariation:Ueq}) to be on-shell. This equates to discarding $\ket{C_i}$ and $\ket{D_i}$, and to require the physical field $\ket{\phi_i}$ to be transverse and to satisfy the free equations of motion defined by the quadratic term in~\eqref{LQ} (i.e. $l^0_i\ket{\phi_i}=l_i\ket{\phi_i}=0$). The cubic interaction between three fields $\ket{\phi_i}$ is defined by the part of a cubic vertex $V$ that contains only $\alpha^{\dagger}_{\mu, i}$ and $\xi_i^\dagger$ oscillators. For contributions coming from the nonlinear part of the gauge transformations, ghost parts of the cubic vertex do contribute as well. Specifically, it will be useful to define $V_i$ as the ghost-free part of the cubic vertex which appears multiplied by the ghost combination $(b^0_{i+1}-b^0_{i+2})c^\dagger_i$. In practice this is equivalent to
	\begin{equation}
		V_i=\frac{\partial V}{\partial\mathcal{Z}}\mathcal{Z}_i+\frac{\partial V}{\partial\mathcal{Q}}\mathcal{Q}_i+\sum_{k=1}^3\frac{\partial V}{\partial\mathcal{K}_k}\delta_k^i\ ,
	\end{equation}
	where
	\begin{equation}
		\begin{split}
		\mathcal{Q}_1&=-\frac{\xi_2^\dagger}{2m}\ ;\quad\quad\,\mathcal{Q}_2=\frac{\xi_1^\dagger}{2m}\ ;\\
		\mathcal{Z}_1&=-\alpha_2^\dagger\cdot\alpha_3^\dagger\ ;\quad%
		\mathcal{Z}_2=-\alpha_3^\dagger\cdot\alpha_1^\dagger\ ;\quad%
		\mathcal{Z}_3=-\alpha_1^\dagger\cdot\alpha_2^\dagger+\xi_1^\dagger\xi_2^\dagger\ .
		\end{split}
	\end{equation}
	The contribution of all $\bra{\phi_i}Q_i$ in e.g.,~\eqref{eq:QuarticVariation:Weq} vanish, since $\bra{\varphi_i}Q_i$ vanishes on-shell. This leaves us with
	\begin{align}
		\begin{split}
			0&=\int dc^0_1dc^0_2dc^0_3dc^0_4\bra{\Lambda_1}\bra{\phi_2}\bra{\phi_3}\bra{\phi_4}\left(Q_1\ket{W}+2\ket{V:_1V}-2\ket{V:_2V}+\frac{g_2}{g}\ket{\Vgr:_3V}\right)\\
			&=\bra{\lambda_1}\bra{\varphi_2}\bra{\varphi_3}\bra{\varphi_4}\bigg(l_1W-2V_1(\bar{1},1',\bar{4})V(\bar{1}',\bar{2},\bar{3})-2V(2',\bar{2},\bar{4})V_1(\bar{1},\bar{2}',\bar{3})\\
            &\hspace{4.75cm}+\frac{g_2}{g}\Vgr(\bar{3},\bar{4},5)V_1(\bar{1},\bar{2},\bar{5})\bigg)\ket{0}\ ;
		\end{split}\\
		\begin{split}
			0&=\int dc^0_1dc^0_2dc^0_3dc^0_4\bra{\phi_1}\bra{\phi_2}\bra{\Lambda_3}\bra{\phi_4}\left(Q_3\ket{W}+2\ket{V:_1V}-2\ket{V:_2V}+\frac{g_2}{g}\ket{\Vgr:_3V}\right)\\
			&=\bra{\varphi_1}\bra{\varphi_2}\bra{\lambda_3}\bra{\varphi_4}\bigg(l_3W-2V(\bar{1},1',\bar{4})V_3(\bar{1}',\bar{2},\bar{3})+2V(2',\bar{2},\bar{4})V_3(\bar{1},\bar{2}',\bar{3})\\
            &\hspace{4.75cm}+\frac{g_2}{g}\Vgri{3}(\bar{3},\bar{4},5)V(\bar{1},\bar{2},\bar{5})\bigg)\ket{0}
		\end{split}
	\end{align}
	Recalling that $V(\bar{2},\bar{1},\bar{3})=\pm V(\bar{1},\bar{2},\bar{3})$ for any vertex constructed of invariants~\eqref{eq:brstcubic:K1}-\eqref{eq:brstcubic:cZ} and applying the techniques of Appendix~\ref{sec:bravsket}, one can rewrite the equation above in a symmetrical form
	\begin{align}
		l_1W&=\left(2V(2',\bar{2},\bar{4})V_1(\bar{1},\bar{2}',\bar{3})+(3\leftrightarrow4)\right)-\frac{g_2}{g}\Vgr(\bar{3},\bar{4},5)V_1(\bar{1},\bar{2},\bar{5})\ \label{o1} ,\\
		l_3W&=\left(2V(\bar{1},1',\bar{4})V_3(\bar{1}',\bar{2},\bar{3})+(1\leftrightarrow2)\right)-\frac{g_2}{g}\Vgri{3}(\bar{3},\bar{4},5)V(\bar{1},\bar{2},\bar{5})\ \label{o2}.
	\end{align}
    Repeating the same procedure for $U$ and $X$ vertices one arrives at
    \begin{align}
         l_1U&=-\Big(V(\bar{1}',\bar{2}',3)V_1(\bar{1},\bar{2},\bar{3})+(2\leftrightarrow 2')\Big) \label{o3}\ ,\\
         l_3X&=-\frac{1}{4}\Big(\Vgr(\bar{3}',\bar{4}',5)\Vgri{3}(\bar{3},\bar{4},\bar{5})+(4\leftrightarrow 4')\Big)\ .\label{o4}
    \end{align}
    The equations \p{o1}--\p{o4} together with reality conditions~\eqref{eq:aqi:RealCond} are defining equations for the on-shell quartic vertices.
  
    \subsection{Example: On-shell Quartic for 1-1-2 Cubic Vertex and Fixing Cubic Coupling}\label{sec:112}

    In this subsection we give a solution to the on-shell equations \p{o1}--\p{o4} when the cubic vertex describes interactions between two massive vector fields and the graviton, i.e.
	\begin{equation} \label{Z112}
		\ket{V}=\mathcal{Z}{\cal K}^{(3)}c_1^0c_2^0c_3^0\ket{0}\ .
	\end{equation}
    Note that the existence of a local solution for $W$ requires the right-hand side of~\eqref{o2} to be at least linear in the momentum $p_3$, since any action of $l_3$ on $W$ will always generate a factor of $p_3$ (up to momentum conservation; e.g., $p_1\cdot p_4=p_2\cdot p_3$ is considered proportional to $p_3$). Computing the right-hand side
    of equation \p{o2}
     we see that 
     it is proportional to $p_3$ if we choose a specific value for the ratio of coupling constants
    \begin{equation}
        \frac{g_2}{g} = 1\ .
    \end{equation}
    Under this choice, we can compute the local $W$ vertex 
    \begin{equation}
    \begin{split}
        W(\bar{1},\bar{2},\bar{3},\bar{4})=\bigg(&\ 8\left(-\A{1}{2} + \xi_1^\dagger\xi_2^\dagger\right)\A{3}{4}\Pa{2}{3}\Pa{1}{4}\\
        +&\ 8\A{3}{4}\A{1}{3} \left(\Pa{3}{4}\Pa{4}{2}+\Pa{1}{4}\Pa{3}{2}-\Pa{2}{4}\Pa{3}{2}\right)\\
        -&\ \A{3}{4}\A{3}{4}\left(4\Pa{3}{1}\Pa{4}{2} + p_3\cdot p_4 (-\A{1}{2} +    \xi_1^\dagger\xi_2^\dagger)\right)\\
        -&\ 4\A{1}{4}\A{2}{3}\left(\Pa{2}{3}\Pa{1}{4}- 2\Pa{1}{3}\Pa{2}{4} - 2p_1\cdot p_3 \A{3}{4}\right)\\
        +&\ (1\leftrightarrow2)\bigg)+(3\leftrightarrow4)
    \end{split}
    \end{equation}
    We can also find the solution for $U$ and $X$ in a straightforward manner. Since $X$ is simply the vertex of four massless spin-1 particles, the results and methods of~\cite{Taronna:2011kt} apply unchanged. A solution for $U$ is given in  Appendix \ref{Q112}.
    
    \section{Conclusion and Outlook}

    In this paper, we developed a systematic procedure
    for finding quartic interaction vertices
    for a massless and massive bosonic field with arbitrary spins.
    While the general formalism is valid for any integer spin massless field, we restricted our study to the massless spin being either
    $2$ (graviton) or $1$ (a vector field), bearing in mind potential applications to black-hole binaries. The number of space-time dimensions was also left arbitrary.

    We presented a general Lagrangian including these quartic vertices and derived consistency equations that relate them to cubic vertices. We then gave a general (non-local) formal solution to those equations valid for any spin. In a low spin example we show that local solutions can also be constructed. The existence of such local solutions appears to depend on the precise form and combination of cubic vertices present in the Lagrangian. To find an explicit and \emph{local} off-shell quartic extension to the higher-spin parts of the  minimal coupling cubic vertices relevant to black-hole binaries remains an open problem. Alternatively, it would be of interest if one can derive general conditions on the cubic vertices that allow for the existence of local quartic vertices. The methods we laid out for computing the vertices by hand should be straightforward to improve upon numerically, allowing for the computation of higher spin results.
    
    Finally, we present on-shell versions of our results in section \ref{seconshell}. This restriction considerably 
    shortened the equations, and allowed us to compute a (still lengthy) result for a massive vector field interacting with a graviton. Since we need to impose symmetry between different auxillary Fock spaces in the formalism by hand, applying the spinor - helicity formalism \cite{Dittmaier:1998nn}-\cite{Conde:2016izb} (when the spacetime equals four) could potentially lead to further simplification. We hope to address this problem in the future.

    \section*{Acknowledgments}  
    We are grateful to  Y. Neiman and E. Skvortsov for useful discussions.
    The work  was supported by the Quantum Gravity Unit of the Okinawa Institute of Science and Technology Graduate University (OIST).

    \renewcommand{\theequation}{A.\arabic{equation}}
    \setcounter{equation}0
    \appendix
    \numberwithin{equation}{section}

    \section{Bra vs. Ket Vectors}\label{sec:bravsket}

    In order to solve the equations for quartic vertices, we need to be able to compare terms written exclusively using ket-vectors (like $\bra{\Lambda_1}\bra{\phi_2}\bra{\phi_3}\bra{\phi_4}\ket{W}$) and terms using both bra- and ket- vectors (like $\bra{V}\ket{\Lambda_{2'}}\ket{\phi_4}\bra{\phi_1}\bra{\phi_3}\ket{V}$).
    We can compare such terms, because at the level of the classical Lagrangian, the component fields of $\ket{\phi_i}$ are either real or related to each other via complex conjugation (e.g., $\varphi_1^{\mu_1\ldots\mu_s}(x)=\varphi_2^{\star\mu_1\ldots\mu_s}(x)$). To explain how we translate such expressions, let us consider the following term including $\ket{\phi_1}$ which we would like to rewrite as
	\begin{equation}
		\label{eq:bravsket:Phi1KetToBra}
		\int dc_1^0\bra{V(1,2',3)}\ket{\phi_1}=\int dc_2^0\bra{0}_{2'3}\otimes\bra{\phi_2} O(\bar{2},2',3)\ket{0}_2\ ,
	\end{equation}
	for some operators $O$. Here, we have made explicit the oscillator dependence of the cubic vertex as was done for contractions; e.g., in~\eqref{eq:dW:ContrDef1}. The massless case $\ket{\phi_3}$ is analogous, but less complicated since the component fields of $\ket{\phi_3}$ are real.
	
    We recall that the vector $\ket{\phi_1}$ can be decomposed into three ghost-free component fields~\eqref{eq:brstcubic:triplet}. We start the discussion of how to turn ket-vectors into bra-vectors with the component field $\ket{\varphi_1}$. We have
	\begin{equation}
		\begin{split}
			\bra{V(1,2',3)}\ket{\varphi_1}=\frac{1}{s!}\sum_{k=0}^s\bra{0}c_3^0c_{2'}^0c_1^0V^\dagger(1,2',3)\varphi_1^{\mu_1\ldots\mu_{s-k}}(x)\alpha_{1\mu_1}^\dagger\ldots\alpha_{1\mu_{s-k}}^\dagger\left(\xi^\dagger_1\right)^k\ket{0}_1
		\end{split}\ .
	\end{equation}
	As a first step, we commute $V^\dagger$ and $\varphi_1^{\mu_1\ldots\mu_s}(x)$. This requires us to partially integrate any $\partial_{1,\mu}$ and therefore flip the sign of the momentum $p_1$. Furthermore, using the fact that $\varphi_1^{\mu_1\ldots\mu_s}(x)=\varphi_2^{\star\mu_1\ldots\mu_s}(x)$ and making the momentum dependence of $V$ explicit, we can write
	\bea
		\bra{V(1,2',3)}\ket{\varphi_1}&=&\frac{1}{s!}\sum_{k=0}^s\bra{0}\varphi_2^{\star\mu_1\ldots\mu_{s-k}}(x)c_3^0c_{2'}^0c_1^0V^\dagger(1,2',3,-p_1,p_{2'},p_3)
        \\ \nonumber
        &\times&\alpha_{1\mu_1}^\dagger\ldots\alpha_{1\mu_{s-k}}^\dagger\left(\xi^\dagger_1\right)^k\ket{0}_1 \, .
	\eea
	Note that the above sign change in $p_1$ is in addition to any sign-flip coming from performing the hermitian conjugation of $V$. Next, we exchange $\alpha_1$ and $\xi_1$ in $V^\dagger$ with $\alpha^\dagger_1$ and $\xi^\dagger_1$ coming from $\ket{\varphi_1}$. We have for any totally symmetric tensors $f$ and $g$
	\begin{equation}
		\bra{0}f_{\mu_1\ldots\mu_s}\alpha^{\mu_1}\ldots\alpha^{\mu_s}\alpha^\dagger_{\nu_1}\ldots\alpha^\dagger_{\nu_s}g^{\nu_1\ldots\nu_s}\ket{0}=\bra{0}f^{\nu_1\ldots\nu_s}\alpha^{\mu_1}\ldots\alpha^{\mu_s}\alpha^\dagger_{\nu_1}\ldots\alpha^\dagger_{\nu_s}g_{\mu_1\ldots\mu_s}\ket{0}\ ,
	\end{equation}
	which leads to
	\bea
		\bra{V(1,2',3)}\ket{\varphi_1}&=&\frac{1}{s!}\sum_{k=0}^s\bra{0}\varphi_2^{\star\mu_1\ldots\mu_{s-k}}(x)\alpha_{1\mu_1}\ldots\alpha_{1\mu_{s-k}}\left(\xi_1\right)^k \\ \nonumber
        &\times &c_3^0c_{2'}^0c_1^0V^\dagger(\bar{1},2',3,-p_1,p_{2'},p_3)\ket{0}_1 .
	\eea
	Finally, relabeling the Fock-space $1$ to $2$, we arrive at 
	\begin{equation}
        \bra{V(1,2',3)}\ket{\varphi_1}=\bra{0}_{2'3}\otimes\bra{\varphi_2} c_3^0c_{2'}^0c_2^0V^\dagger(\bar{2},2',3,-p_2,p_{2'},p_3)\ket{0}_2\ .
	\end{equation}
	We have arrived precisely at an expression of shape~\eqref{eq:bravsket:Phi1KetToBra}. Let us now consider the effect of ghosts. Starting with the term including $\ket{D_1}$, (see \eqref{eq:brstcubic:triplet}) we have
	\begin{equation}
		\bra{V(1,2',3)}c_1^\dagger b_1^\dagger\ket{D_1}=\bra{0}c_3^0c_{2'}^0c_1^0V^\dagger(1,2',3)c_1^\dagger b_1^\dagger\ket{D_1}\ .
	\end{equation}
	This expression is non-vanishing only for  terms in $V^\dagger$ that contain exactly the ghosts $c_1b_1$ in the first Fock-space. While in principle $b_1^0c_1b_1$ is also possible, it will not contribute since 
    we integrate over $dc_1^0$ in~\eqref{eq:bravsket:Phi1KetToBra}. We can now permute the rest of $V^\dagger$ such that to replace $c_1b_1$ with $c^\dagger_1b^\dagger_1$ in $V^\dagger$. Permuting the remaining $c_1b_1$ all the way to the left, and swapping bosonic oscillators from creation to annihilation as before, we obtain
	\begin{equation}
		\int dc_1^0\bra{V(1,2',3)}c_1^\dagger b_1^\dagger\ket{D_1}=\int dc_2^0\bra{0}_{2'3}\otimes\bra{D_2}c_2b_2\  c_3^0c_{2'}^0c_2^0V^\dagger(\bar{2},2',3,-p_2,p_{2'},p_3)\ket{0}_2\ .
	\end{equation}
	Finally, let us consider the component $\ket{C_1}$
    in \p{eq:brstcubic:triplet},
	\begin{equation}
		\bra{V(1,2',3)}c_1^0 b_1^\dagger\ket{C_1}=\bra{0}c_3^0c_{2'}^0c_1^0V^\dagger(1,2',3)c_1^0 b_1^\dagger\ket{C_1}\ .
	\end{equation}
	This time, any parts in $V^\dagger$ that can contribute to this expression have to include $b_1^0c_1$. We anti-commute $c_1$ coming from $V^\dagger$ past the right $c_1^0$ to cancel the $b_1^\dagger$, producing one minus sign. We can then reinsert $1=b_1 c_1^\dagger$ on the left of $c_3^0$. Next we anti-commute $c_1^\dagger$ back into the position where $c_1$ was before. Since $b_1^0$ is real, this is all we need to swap daggered for undaggered operators in the ghost parts. Lastly, we again swap the bosonic parts as before, which leaves us with
	\begin{equation}
		\int dc_1^0\bra{V(1,2',3)}c_1^0 b_1^\dagger\ket{C_1}=-\int dc_2^0\bra{0}_{2'3}\otimes\bra{C_2}b_2\  c_3^0c_{2'}^0c_2^0V^\dagger(\bar{2},2',3,-p_2,p_{2'},p_3)c_2^0\ket{0}_2\ .
	\end{equation}
	We can now collect all three pieces. Performing some additional commutations of ghost oscillators to bring them in a common form, we find
	\begin{equation}
		\begin{split}
		\int dc_1^0\bra{V(1,2',3)}\ket{\phi_1}&=\\
		\int dc_2^0\bra{0}_{2'3}\otimes&\big(\bra{\varphi_2}-\bra{C_2}b_2c_2^0-\bra{D_2}b_2c_2\big)c_3^0c_{2'}^0V^\dagger(\bar{2},2',3,-p_2,p_{2'},p_3)c_2^0\ket{0}_2\ .
		\end{split}
	\end{equation}
    The only difference to $\bra{\phi_2}$ is a sign-flip in the ghost parts. We can account for this sign by changing the sign of every term containing $c_2^\dagger$ in $V^\dagger$. We arrive at the result
	\begin{equation}\label{eq:bravsket:KetToBra1}
		\int dc_1^0\bra{V(1,2',3)}\ket{\phi_1}=\int dc_2^0
		\bra{0}_{2'3}\otimes\bra{\phi_2}c_3^0c_{2'}^0\left(V^\dagger(\bar{2},2',3,-p_2,p_{2'},p_3)\big\vert_{c_2^\dagger\rightarrow-c_2^\dagger}\right)c_2^0\ket{0}_2\ ;
	\end{equation}
    Arguments like the above can be used to convert any kind of ket to bra operation that are necessary in this paper. We collect below some of the most important identities:
    \begin{equation}
    \begin{split}
        \int dc_1^0dc_3^0\bra{V(1,2',3)}\ket{\phi_1}\ket{\phi_3}&=\int dc_2^0dc_3^0
		\bra{\phi_2}\bra{\phi_3}\bra{0}_{2'}c_{2'}^0V(\bar{2},2',\bar{3})\big\vert_{c_{2'}^\dagger\rightarrow-c_{2'}^\dagger}c_2^0c_3^0\ket{0}_{23}\ ;\\
        \int dc_1^0dc_2^0\bra{V(1,2,3)}\ket{\phi_1}\ket{\phi_2}&=\int dc_2^0dc_1^0
		\bra{\phi_1}\bra{\phi_2}\bra{0}_{3}c_{3}^0V(\bar{2},\bar{1},3)\big\vert_{c_{3}^\dagger\rightarrow-c_{3}^\dagger}c_1^0c_2^0\ket{0}_{23}\ ;\\
        \int dc_1^0dc_3^0\bra{V(1,2',3)}\ket{\Lambda_1}\ket{\phi_3}&=-\int dc_2^0dc_3^0
		\bra{\Lambda_2}\bra{\phi_3}\bra{0}_{2'}c_{2'}^0V(\bar{2},2',\bar{3})\big\vert_{c_{2'}^\dagger\rightarrow-c_{2'}^\dagger}c_2^0c_3^0\ket{0}_{23}\ ;\\
        \int dc_1^0dc_3^0\bra{V(1,2',3)}\ket{\Lambda_1}\ket{\Xi_3}&=\int dc_2^0dc_3^0
		\bra{\Lambda_2}\bra{\Xi_3}\bra{0}_{2'}c_{2'}^0V(\bar{2},2',\bar{3})\big\vert_{c_{2'}^\dagger\rightarrow-c_{2'}^\dagger}c_2^0c_3^0\ket{0}_{23}\ .
    \end{split}
    \end{equation}
    We can use expressions such as these in order to arrive at the contractions~\eqref{eq:dW:ContrDef1}-~\eqref{eq:dW:ContrDef3}. For example, one finds
    \begin{equation}
    \begin{split}
		&\int dc^0_{1'}dc_{2}^0dc_{3}^0dc^0_4\bra{V(1',2,4)}\ket{\phi_{4}}\ket{\Xi_{1'}}\bra{\Lambda_3}\ket{V(\bar{1},\bar{2},\bar{3})}=\\
		-&\int dc_{2'}^0dc_{2}^0dc^0_{3}dc^0_{4} \bra{\Xi_{2'}}\bra{\phi_{4}}\bra{0}_{2}c_{2}^0\times\\
        &\hspace{1.5cm}\left(V(2,\bar{2}',\bar{4},-p_{2'}-p_4,p_{2'},p_4)\big\vert_{c_{2}\rightarrow -c_{2}}\right)c_{2'}^0c_{4}^0\ket{0}_{2'4}\bra{\Lambda_3}\ket{V(\bar{1},\bar{2},\bar{3})}=\\
        &\int dc_2^0dc_{2'}^0dc^0_{3}dc^0_{4} \bra{\Xi_{2}}\bra{\Lambda_3}\bra{\phi_4}\bra{0}_{2'}c_{2'}^0\times\\
        &\hspace{1.5cm}\left(V(2',\bar{2},\bar{4},-p_{2}-p_4,p_{2},p_4)\big\vert_{c_{2'}\rightarrow -c_{2'}}\right)c_{2}^0c_{4}^0V(\bar{1},\bar{2'},\bar{3})c_1^0c_{2'}^0c_3^0\ket{0}=\\
        -&\int dc_2^0dc^0_{3}dc^0_{4} \bra{\Xi_{2}}\bra{\Lambda_3}\bra{\phi_4}\ket{V:_2V}\ .
    \end{split}
	\end{equation}
    In the first equality we applied momentum-conservation in the first cubic vertex to eliminate $p_2=-p_{1'}-p_4$ and then applied the techniques of this section; in the second we relabeled $2\leftrightarrow 2'$ and anti-commuted $\bra{\Lambda_3}$ to the left; in the third equality we applied definition~\eqref{eq:dW:ContrDef2}.
    
    \section{Derivation of the Quartic Vertex Consistency Equations}\label{sec:QuarticVariation}
    \subsection{Gauge Invariance of the Quartic Lagrangian}
    In this section we provide additional computational details for the derivation of equations~\eqref{eq:dW:Weq}-\eqref{eq:dW:Xeq} from the gauge variation of the Lagrangian~\eqref{LQ}. Varying with respect to the massive field, we obtain the following terms proportional to gauge parameter $\ket{\Lambda_1}$
	\begin{equation} \label{l11}
		\begin{split}
			\delta_{\Lambda_1}\mathcal{L}=&\quad\;4g\int dc^0_1dc^0_2dc^0_3\bra{\Lambda_1}\bra{\phi_2}\bra{\phi_3}(Q_1+Q_2+Q_3)\ket{V}\\
			&-8g^2\int dc^0_1dc^0_2dc^0_4\bra{V(1,2,4)}\ket{\phi_{4}}\ket{\phi_{1}}\left(\int dc^0_{3}dc^0_{1'}\bra{\phi_{3}}\bra{\Lambda_{1'}}\ket{V(\bar{1}',\bar{2},\bar{3})}\right)\\
			&-4g^2\int dc^0_1dc^0_2dc^0_3\bra{V(1,2,3)}\ket{\phi_{2}}\ket{\phi_{1}}\left(\int dc^0_{1'}dc^0_{2'}\bra{\Lambda_{1'}}\bra{\phi_{2'}}\ket{V(\bar{1'},\bar{2'},\bar{3})}\right)\\
            &-2gg_2\int dc^0_3dc^0_4dc^0_5\bra{\Vgr(3,4,5)}\ket{\phi_{3}}\ket{\phi_{4}}\left(\int dc^0_{1}dc^0_{2}\bra{\Lambda_{1}}\bra{\phi_{2}}\ket{V(\bar{1},\bar{2},\bar{5})}\right)\\
			&+2g^2\int dc^0_1dc^0_2dc^0_3dc^0_4\bra{\Lambda_1}\bra{\phi_2}\bra{\phi_3}\bra{\phi_4}(Q_1+Q_2+Q_3+Q_4))\ket{W}\\
			&+2g^2\int dc_1^0dc^0_2dc_{1'}^0dc_{2'}^0\bra{\Lambda_1}\bra{\phi_2}\bra{\phi_{1'}}\bra{\phi_{2'}}(Q_1+Q_2+Q_{1'}+Q_{2'})\ket{U}\ .
		\end{split}
	\end{equation}
    The remaining terms from the variation of the massive field are proportional to $\ket{\Lambda_2}$ and are simply the complex conjugates of the above. We see that the term that is first order in the couplings indeed vanishes under~\eqref{eq:brstcubic:BRSTInvarianceCubic}. For the quartic vertices, we can collect terms with the same fields to separate the $U$ and $W$ equations. This leaves us with
    \begin{align}
        \begin{split}\label{eq:QuarticVariation:Weq}
            &\int dc^0_1dc^0_2dc^0_3dc^0_4\bra{\Lambda_1}\bra{\phi_2}\bra{\phi_3}\bra{\phi_4}(Q_1+Q_2+Q_3+Q_4))\ket{W}\\
			&=4\int dc^0_1dc^0_2dc^0_4\bra{V(1,2,4)}\ket{\phi_{4}}\ket{\phi_{1}}\left(\int dc^0_{3}dc^0_{1'}\bra{\phi_{3}}\bra{\Lambda_{1'}}\ket{V(\bar{1}',\bar{2},\bar{3})}\right)\\
            &\quad\,+\frac{g_2}{g}\int dc^0_3dc^0_4dc^0_5\bra{\Vgr(3,4,5)}\ket{\phi_{3}}\ket{\phi_{4}}\left(\int dc^0_{1}dc^0_{2}\bra{\Lambda_{1}}\bra{\phi_{2}}\ket{V(\bar{1},\bar{2},\bar{5})}\right)\\
            &=\int dc^0_1dc^0_2dc^0_3dc^0_4\bra{\Lambda_1}\bra{\phi_2}\bra{\phi_3}\bra{\phi_4}\left(4\ket{V:_2V}-\frac{g_2}{g}\ket{\Vgr:_3V}\right)\ ,
        \end{split}\\
        \begin{split}\label{eq:QuarticVariation:Ueq}
            &\int dc_1^0dc^0_2dc_{1'}^0dc_{2'}^0\bra{\Lambda_1}\bra{\phi_2}\bra{\phi_{1'}}\bra{\phi_{2'}}(Q_1+Q_2+Q_{1'}+Q_{2'})\ket{U}\\
            &=2\int dc^0_1dc^0_2dc^0_3\bra{V(1,2,3)}\ket{\phi_{2}}\ket{\phi_{1}}\left(\int dc^0_{1'}dc^0_{2'}\bra{\Lambda_{1'}}\bra{\phi_{2'}}\ket{V(\bar{1'},\bar{2'},\bar{3})}\right)\\
            &=-2\int dc_1^0dc^0_2dc_{1'}^0dc_{2'}^0\bra{\Lambda_1}\bra{\phi_2}\bra{\phi_{1'}}\bra{\phi_{2'}}\ket{V:_3V}
        \end{split}
    \end{align}
    where we have used definitions~\eqref{eq:dW:ContrDef1}-\eqref{eq:dW:ContrDef3} as well as methods from Appendix~\eqref{sec:bravsket}. The first equation~\eqref{eq:QuarticVariation:Weq} can be rearranged, exploiting the symmetry under relabeling oscillators $3\leftrightarrow4$. Using symmetries~\eqref{eq:dW:contractionSymmetry} this results in~\eqref{eq:dW:Weq} as expected. For the $U$-equation~\eqref{eq:QuarticVariation:Ueq} we use the freedom of relabeling $2\leftrightarrow2'$ which, together with~\eqref{eq:dW:contractionSymmetry}, results in~\eqref{eq:dW:Ueq}. On the other hand, variation with respect to the massless field gives us the same cubic condition on $V$, the $W$-equation~\eqref{eq:dW:Weq}, plus the additional terms
    \begin{equation}
		\begin{split}
			\delta_{\Lambda_3}\mathcal{L}&=O(g)+W\text{-terms }+\\
            &+2g_2\int dc_3^0dc_4^0dc_5^0\bra{\Lambda_3}\bra{\phi_4}\bra{\phi_5}(Q_3+Q_4+Q_5)\ket{\Vgr}\\
            &-2g_2^2\int dc^0_{3'}dc^0_{4'}dc^0_5\bra{\Vgr(3',4',5)}\ket{\phi_{3'}}\ket{\phi_{4'}}\left(\int dc^0_{3}dc^0_{4}\bra{\Lambda_{3}}\bra{\phi_{4}}\ket{\Vgr(\bar{3},\bar{4},\bar{5})}\right)\\
			&+4g^2\int dc_3^0dc^0_4dc_{3'}^0dc^0_{4'}\bra{\Lambda_3}\bra{\phi_4}\bra{\phi_{3'}}\bra{\phi_{4'}}(Q_3+Q_4+Q_{3'}+Q_{4'})\ket{X}
		\end{split}
	\end{equation}
    The term linear in $g_2$ again vanishes due to~\eqref{eq:brstcubic:BRSTInvarianceCubic}. The contraction term is analogous to the case of $U$ from above and can be simplified using~\eqref{eq:dW:ContrDef3} which, using relabeling symmetries of massless fields $4\leftrightarrow4'$ as well as symmetries~\eqref{eq:dW:contractionSymmetry}, results in~\eqref{eq:dW:Xeq}.

	\subsection{Closure of the Algebra of Gauge Transformations}\label{sec:dY}

    In addition to requiring gauge invariance of the Lagrangian, we have to verify that the algebra of gauge transformations closes, i.e. for two gauge parameters $\Lambda_i$ and $\Xi_i$
	\begin{equation}\label{eq:dY:gaugeclosure}
        \left[\delta_{\Xi},\delta_{\Lambda}\right]\ket{\phi_i}\approx\delta_{f_i(\Lambda,\Xi)}\ket{\phi_i} \ ,
	\end{equation}
    where the $\approx$ indicate that we only require this to hold for $\ket{\phi_i}$ that are solutions to the equation of motion. The closure of the algebra at the level of cubic interactions is equivalent to the requirement of BRST invariance of the cubic vertex~\eqref{eq:brstcubic:BRSTInvarianceCubic} (see \cite{Neveu:1986mv} for analogous consideration in the Bosonic Open String Field Theory). It fixes
    \begin{equation} \label{iden}
    \begin{split}
        \ket{f_i(\Lambda,\Xi)}=(2-\delta_i^3)g&\int dc^0_{i+1}dc^0_{i+2}\left(\bra{\Xi_{i+1}}\bra{\Lambda_{i+2}}-\bra{\Lambda_{i+1}}\bra{\Xi_{i+2}}\right)\ket{V}\\
        +\delta_i^3\;g_2&\int dc^0_{4}dc^0_{5}\bra{\Xi_{4}}\bra{\Lambda_{5}}\ket{\Vgr}\ ,
    \end{split}
    \end{equation}
    for $i=1,\ 2,\ 3$ at first order in the coupling constants. We need to show that by imposing the conditions coming from the closure of the gauge algebra at the quadratic order in the couplings we recover conditions~\eqref{eq:dW:Weq}-\eqref{eq:dW:Xeq} on the quartic vertices. We will start by explicitly computing~\eqref{eq:dY:gaugeclosure} for $\ket{\phi_1}$. Applying the gauge transformations~\eqref{eq:offshell_gaugetransf1}-\eqref{eq:offshell_gaugetransf} twice, we obtain
	\begin{equation}
			\begin{split}
			\delta_{\Xi}\delta_{\Lambda}\ket{\phi_1}&=2g\int dc^0_{2}dc^0_{3}\left(-\bra{\Xi_{2}}\bra{\Lambda_{3}}Q_2+\bra{\Lambda_{2}}\bra{\Xi_{3}}Q_3\right)\ket{V}\\
            &-4g^2\int dc^0_{2}dc^0_{3}dc_4^0\left(\bra{\Xi_2}\bra{\Lambda_{3}}\bra{\phi_4}-\bra{\phi_2}\bra{\Lambda_3}\bra{\Xi_{4}}\right)\ket{V:_2V}\\
            &-2gg_2\int dc_2^0dc_3^0dc_4^0\bra{\Lambda_2}\bra{\phi_3}\bra{\Xi_4}\ket{\Vgr:_3V}\\
			&-2g^2\int dc^0_{2}dc^0_{3}dc^0_{4}\left(\bra{\Lambda_{2}}\bra{\Xi_{3}}\bra{\phi_{4}}Q_3+\bra{\phi_{2}}\bra{\Lambda_{3}}\bra{\Xi_{4}}Q_4\right.\\
            &\hspace{3.1cm}\left.-\bra{\Xi_{2}}\bra{\Lambda_{3}}\bra{\phi_{4}}Q_2\right)\ket{W}\\
            &-2g^2\int dc_{2}^0dc_{1'}^0dc^0_{2'}\left(\bra{\Lambda_2}\bra{\phi_{1'}}\bra{\Xi_{2'}}+\bra{\Lambda_2}\bra{\Xi_{1'}}\bra{\phi_{2'}}\right)\ket{V:_3V}\\
			&-2g^2\int dc_2^0dc_{1'}^0dc_{2'}^0\left(\bra{\Lambda_2}\bra{\Xi_{1'}}\bra{\phi_{2'}}Q_{1'}+\bra{\Lambda_2}\bra{\phi_{1'}}\bra{\Xi_{2'}}Q_{2'}\right.\\
            &\hspace{3.3cm}\left.-\bra{\Xi_2}\bra{\Lambda_{1'}}\bra{\phi_{2'}}Q_2\right)\ket{U}\ .
		\end{split}
	\end{equation}
    After forming the commutator this has to equal the RHS
    \begin{equation}
        \begin{split}
            \delta_{f_i(\Lambda,\Xi)}\ket{\phi_1}=2g&\int dc_2^0dc_3^0\left(\bra{\Xi_2}\bra{\Lambda_3}-\bra{\Lambda_2}\bra{\Xi_3}\right)Q_1\ket{V}\\
            +2g^2&\int dc_2^0dc_{1'}^0dc_{2'}^0\left(\bra{\phi_2}\bra{\Lambda_{1'}}\bra{\Xi_{2'}}-\bra{\phi_2}\bra{\Xi_{1'}}\bra{\Lambda_{2'}}\right)\ket{V:_3V}\\
            +2gg_2&\int dc_{2}^0dc_{3}^0dc_4^0\bra{\phi_2}\bra{\Lambda_{3}}\bra{\Xi_{4}}\ket{\Vgr:_3V}\\
            -4g^2&\int dc_2^0dc_3^0dc_4^0\left(\bra{\Lambda_{2}}\bra{\phi_3}\bra{\Xi_{4}}-\bra{\Xi_{2}}\bra{\phi_3}\bra{\Lambda_{4}}\right)\ket{V:_2V}.
        \end{split}
    \end{equation}
    Combining the two, we see that the linear order in coupling is indeed satisfied by condition~\eqref{eq:brstcubic:BRSTInvarianceCubic}. At quartic order we get for $W$ and $U$
	\begin{align}
        \begin{split}
            0=&\int dc^0_{2}dc^0_{3}dc^0_{4}\bra{\Lambda_{2}}\bra{\Xi_{3}}\bra{\phi_{4}}\times\\
            &\left((Q_2+Q_3)\ket{W}+2\ket{V:_1V}-2\ket{V:_2V}+\frac{g_2}{g} \ket{\Vgr:_3V}\right)
        \end{split}\\
        \begin{split}
            0=&\int dc^0_{2}dc^0_{3}dc^0_{4}\bra{\phi_{2}}\bra{\Lambda_{3}}\bra{\Xi_{4}}\times\\
            &\left((Q_3+Q_4)\ket{W}+2\ket{V:_1V}-2\ket{V:_2V}+\frac{g_2}{g} \ket{\Vgr:_3V}\right)
        \end{split}\\
        \begin{split}
            0=&\int dc_2^0dc_{1'}^0dc_{2'}^0\bra{\Lambda_2}\bra{\Xi_{1'}}\bra{\phi_{2'}}\left((Q_{1'}+Q_2)\ket{U}+\ket{V:_3V}-\ket{V:_3V}\big\vert_{2\leftrightarrow2'}\right)
        \end{split}\\
        \begin{split}
            0=&\int dc_2^0dc_{1'}^0dc_{2'}^0\bra{\Lambda_2}\bra{\phi_{1'}}\bra{\Xi_{2'}}\left((Q_2+Q_{2'})\ket{U}+\ket{V:_3V}-\ket{V:_3V}\big\vert_{2\leftrightarrow2'}\right).
        \end{split}
    \end{align}
    We see that those are nearly the same as conditions~\eqref{eq:dW:Weq} and~\eqref{eq:dW:Ueq}. The only difference is, that some $Q_i$ are missing in front of the quartic vertices. We can compensate for the missing $Q_1$ by fixing contributions to~\eqref{iden} at quadratic order in the coupling:
    \begin{equation}
    \begin{split}
        \ket{f_1(\Lambda,\Xi)}&=O(g,g_2)\\
        &\quad+2g^2\int dc^0_{2}dc^0_{3}dc^0_{4}\left(\bra{\Lambda_2}\bra{\Xi_3}\bra{\phi_4}-\bra{\Xi_2}\bra{\Lambda_3}\bra{\phi_4}+\bra{\phi_2}\bra{\Lambda_3}\bra{\Xi_4}\right)\ket{W}\\
        &\quad+2g^2\int dc^0_2dc^0_{1'}dc^0_{2'}\left(\bra{\Lambda_2}\bra{\Xi_{1'}}\bra{\phi_{2'}}-\bra{\Xi_2}\bra{\Lambda_{1'}}\bra{\phi_{2'}}+\bra{\Lambda_2}\bra{\phi_{1'}}\bra{\Xi_{2'}}\right)\ket{U}
    \end{split}
    \end{equation}
    And similar terms for the other $f_i$ and for $X$. This leaves only one missing $Q_i$, namely
    \begin{equation}
        \int dc_2^0dc_{3}^0dc_{4}^0 \bra{\Lambda_2}\bra{\Xi_{3}}\bra{\phi_{4}}Q_{4}\ket{W}\ .
    \end{equation}
    We can rewrite this piece using the EOM of~\eqref{LQ}, which tells us that~$\int dc_{4}^0 \bra{\phi_{4}}Q_{4}$ is actually first order in the coupling. Since this term enters the closure condition already at order $g^2$, it is in total \emph{third} order in the coupling constants and therefore should be omitted. The argument for the other vertices is analogous. Therefore, we see that the closure of the gauge algebra is indeed guaranteed by  conditions~\eqref{eq:dW:Weq}-\eqref{eq:dW:Xeq}.

    \section{Example: On-shell Quartic Solution for $U$ for the 1-1-2 cubic Vertex} \label{Q112}

    We give here the on-shell solution to the $U$ vertex from section~\ref{sec:112}. One could possibly find this solution using numerical methods, by simply solving for the coefficients of the most general polynomial Ansatz. Here, we chose an analytical approach, where the solution was constructed by integrating pieces of the full equation one at a time. When doing so, it is helpful to start with terms missing $\xi^\dagger_i$ oscillators, and then add any necessary pieces that depend on more and more $\xi$-oscillators. It is likely that a somewhat shorter solution than the one presented here can be found by adding appropriate homogeneous solutions. Since this would involve a large computational effort
    with no apparent gain, we have chosen to simply present the solution at hand. The point of this solution is to show that it exists, is local, and non-vanishing.  
    \begin{equation}
    \begin{split}
        U=\Bigg(\bigg(\Big(&-2\A{1'}{2'}\big(\Pa{1'}{2}\Pa{1'}{1}-2\Pa{2'}{2}\Pa{1'}{1}+2\Pa{1'}{2}\Pa{2}{1}\\
        &\hspace{1.5cm}-\Pa{2'}{2}\Pa{2}{1}-3m\Pa{2'}{2}\xi_1^\dagger -m^2\xi_1^\dagger\xi_2^\dagger\big)\\
        &+\A{2}{2'}\big(2 \Pa{2'}{1}\Pa{2'}{1'}+\Pa{1'}{1}\Pa{1}{1'}- m^2 \xi_1^\dagger\xi_{1'}^\dagger\big)\\ 
        &-\frac{4}{m}\Pa{2'}{1'}\xi_1^\dagger\big(\Pa{1'}{2'}\Pa{1'}{2}+2 \Pa{1'}{2'}\Pa{2'}{2}+3\Pa{2}{2'}\Pa{2'}{2}\big)\\
        &+\xi_1^\dagger\xi_{1'}^\dagger\big(-2\Pa{1'}{2}\Pa{1'}{2'}+5\Pa{2'}{2}\Pa{2}{2'}\big)-\frac{7 m^2}{2}\xi_{1}^\dagger\xi_{1'}^\dagger\xi_{2}^\dagger\xi_{2'}^\dagger\\ 
        &+2\xi_{1'}^\dagger\xi_{2'}^\dagger\big(6\Pa{2'}{2}\Pa{2'}{1}-2\Pa{1'}{2}\Pa{2'}{1}+2\Pa{2'}{2}\Pa{1'}{1}-m\Pa{2'}{1}\xi_2^\dagger-m\Pa{1'}{1}\xi_2^\dagger\big)\\    
        &+\frac{2p_1\cdot p_2}{m^2}\big(-m \A{1'}{2'}\big(m \A{1}{2}+2\Pa{1'}{2}\xi_1^\dagger-2\Pa{1}{2}\xi_1^\dagger\big)+\Pa{1}{2}\xi_1^\dagger\xi_{1'}^\dagger\big(5\Pa{1}{2'}+ 3\Pa{2}{2'}\big)\\
        &\hspace{1.5cm}+2\xi_{1'}^\dagger\xi_{2'}^\dagger\big(3\Pa{1'}{2}\Pa{2'}{1}-\Pa{1}{2}\Pa{2}{1}\big)+3 m \xi_{1}^\dagger\xi_{2}^\dagger\xi_{2'}^\dagger\big(2\Pa{2'}{1'}+\Pa{2}{1'}\big)\big)\\
        &+\frac{p_1\cdot p_{1'}}{m^2}\xi_1^\dagger\big(-4 m \A{1'}{2'}\Pa{1}{2}+2 m^2 \A{2}{2'}\xi_{1'}^\dagger-2\Pa{2'}{2}\Pa{2}{2'}\xi_{1'}^\dagger+ 2\Pa{1}{2}\Pa{2'}{1'}\xi_{2'}^\dagger\\
        &\hspace{1.5cm}-4 \Pa{1'}{2}\Pa{2}{1'}\xi_{2'}^\dagger - 12\Pa{1}{2}\Pa{2}{1'}\xi_{2'}^\dagger+ 34 m\Pa{2}{1'}\xi_{2}^\dagger\xi_{2'}^\dagger- 4 m^2\xi_{2}^\dagger\xi_{1'}^\dagger\xi_{2'}^\dagger\big)\\
        &+\frac{p_1\cdot p_2\ p_1\cdot p_{2'}}{m^3}\xi_{1}^\dagger\xi_{2}^\dagger\big(4 m \A{1'}{2'}- 12 \Pa{2'}{1'}\xi_{2'}^\dagger+\frac{9 m}{2}\xi_{1'}^\dagger\xi_{2'}^\dagger \big)\\
        &+\frac{p_1\cdot p_2\ p_1\cdot p_{1'}}{m^3}\xi_{1}^\dagger\xi_{2}^\dagger \big(4 m \A{1'}{2'}-12\Pa{2'}{1'}\xi_{2'}^\dagger+4\Pa{2}{1'}\xi_{2'}^\dagger\big)\\
        &+\frac{p_1\cdot p_{1'}\ p_1\cdot p_{1'}}{m^3}\xi_{1}^\dagger\xi_{2}^\dagger\xi_{2'}^\dagger\big(-8\Pa{2}{1'}+m\xi_{1'}^\dagger\big)\\
        &+\frac{p_1\cdot p_{1'}}{2m^4}\big(p_1\cdot p_{1'}\ p_1\cdot p_{1'}
     - 11 p_1\cdot p_{2}\ p_1\cdot p_{2'}\big)\xi_{1}^\dagger\xi_{2}^\dagger\xi_{1'}^\dagger\xi_{2'}^\dagger\Big)\\
     &+(1\leftrightarrow1')\bigg)+(2\leftrightarrow2')\Bigg)+(1\leftrightarrow2,1'\leftrightarrow2')\ .       
    \end{split}
    \end{equation}

    
	\bibliographystyle{ytphys}
	\bibliography{MasterLibrary}

\end{document}